\documentclass[showpacs,twocolumn]{revtex4}
\usepackage[dvips]{epsfig}

\newcommand{\be}{\begin{equation}}
\newcommand{\ee}{\end{equation}}

\newcommand{\dvr}{\mbox{div}\,}

\newcommand{\bea}{\begin{eqnarray}}
\newcommand{\eea}{\end{eqnarray}}

\newcommand{\prt}{\partial}

\newcommand{\rgl}{\rangle}
\newcommand{\lgl}{\langle}

\begin{document}

\title{On log--normal distribution on a comb structure}
\author{E. Baskin and A. Iomin }

\affiliation{Department of Physics and Solid State Institute, \\ 
Technion, Haifa, 32000, Israel  }

\begin{abstract}
We study specific properties of particles transport  by
exploring an exact solvable model, a so-called comb structure,
where diffusive transport of particles leads to 
subdiffusion.
A performance of L\'evy -- like process enriches this transport 
phenomenon. It is shown that 
an inhomogeneous convection flow, as a realization of the L\'evy--like 
process, leads to superdiffusion of particles on the 
comb structure. A frontier case of superdiffusion
that corresponds to the exponentially fast transport is studied 
and the log--normal distribution is obtained for this case.

\end{abstract}

\maketitle

Fractional kinetics attracts much attention and various aspects of
these phenomena are reflected in the recent reviews 
\cite{shles93,hillfer,klafter,zaslavsky,skb}.  
It should be underlined that while subdiffusion described in the 
framework of the fractional equations has numerous applications 
\cite{hillfer,klafter,oldham,mont,shlezinger}, there are only few 
examples on 
fractional superdiffusion equations \cite{zaslavsky,grigolini,fogedby} 
related to L\'evy like process in 
dynamical chaos \cite{zaslavsky,grigolini} and consideration of 
L\'evy process for the Langevin equations \cite{fogedby}. As it was 
mentioned in 
\cite{skb} ``the superdiffusion is far from being completely understood''.
Indeed, it is difficult to imagine the real physical process with the 
infinite variance jump length during the finite time (L\'evy flight).
In the Letter we present an example of real physical process which is 
absolutely analogous to the L\'evy flights. In this example we discuss 
a behavior of observable quantities, namely a position of the center
of mass of traveling contaminant and the form of the packet (the tail of 
a distribution function). The physical mechanism of this L\'evy--like
effect is an inhomogeneous convection, and the space--time evolution 
of the contaminant in the presence of the inhomogeneous convection flow
is studied. To emphasize this relation between the inhomogeneous 
convection and the L\'evy process, we consider the transport phenomenon on 
a subdiffusive substrate, {\em e.g.} on a comb structure.

Our objective is to explain some complications and a possible mechanism 
of superdiffusion on the comb structure depicted in Fig. 1. 
It is an analogue of a subdiffusive media where subdiffusion has been
already observed \cite{baskin1}. 
A comb model is known as a toy model for a porous medium  
used for exploration of low dimensional percolation clusters \cite{em1}
and electrophoresis process \cite{baskin2}.
It should be underlined that conditions or changes that are 
necessary to perform in the Liouville equation in order to observe 
superdiffusion in the comb model are important for understanding
the nature of this process from the general point of view. 
We show that a performance of L\'evy--like process along, say 
$x$--direction, is due to the inhomogeneity of the 
corresponding $x$--component of the convection current in the Liouville
equation. 
The diffusion process in such media (modeled by the
comb structure) are anomalously slow with the subdiffusive mean squared 
displacement of the order of $\lgl x^2(t)\rgl\sim t^{\alpha},~\alpha<1$.
There are external forces leading to convection. In general case, the 
velocity of the convection flow is space dependent, {\em i.e.} convection
is inhomogeneous. The question under investigation is how the observable shape 
of the initial packet changes, when the space--time evolution of the 
packet corresponds to the convection flow. It should be 
underlined that this problem arises in a variety of applications
starting from transport of external species (pollution) in water flows 
through porous geological formations \cite{berkowitz,hilfer}, problems of 
diffusion and reactions in porous catalysts \cite{andrade} and fractal
physiology \cite{west,from_h,iomin}. We find  the conditions for the 
inhomogeneous convection velocity, when the observable packet's shape
corresponds to the normal diffusion with $\alpha =1$ for the mean squared 
displacement, or even superdiffusion when $\lgl x^2(t)\rgl\sim 
t^{\alpha},~\alpha>1$.   The inhomogeneity will be taken in the form of a 
power law function $\sim v_sx^s$. In this case, the infinitely long 
flights take  place  during a finite fixed time. We will show that the 
L\'evy process takes place
when $s>0$. Our primary objective is the consideration of a frontier case 
when $s=1$. The corresponding solution of the Liouville equation for the 
distribution function of the transport particles is a  
so--called log--normal distribution.
\begin{figure}
\begin{center}
\epsfxsize=6.6cm
\leavevmode
    \epsffile{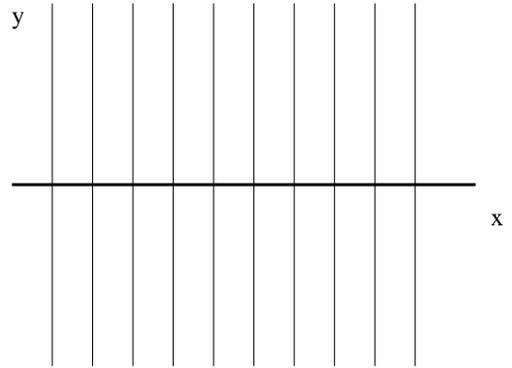}
\caption{The comb structure}
\end{center}
\end{figure}

First, we describe the comb model. A special behavior of the diffusion on 
the comb structure, depicted in 
Fig. 1, is that the displacement in the $x$--direction is possible only 
along the structure axis ($x$-axis, say  at $y=0$). Therefore the diffusion 
coefficient as 
well as a mobility is highly inhomogeneous in the $y$-direction. Namely, 
the diffusion coefficient is $D_{xx}=\tilde{D}\delta(y)$, while the 
diffusion coefficient in the $y$--direction along the teeth is constant 
$D_{yy}=D$. The Liouville equation
\be\label{LE}
\frac{\prt G}{\prt t}+\dvr {\bf j}=0
\ee
describes a random walk on this structure with the distribution function
$G=G(t,x,y)$, and the current 
${\bf j}=(-\tilde{D}\delta(y)\frac{\prt G}{\prt x},\, -D
\frac{\prt G}{\prt y})$. Therefore, it
corresponds to the following Fokker--Planck equation  
\be\label{comb1}
\frac{\prt G}{\prt t}-\tilde{D}\delta(y)\frac{\prt^2G}{\prt x^2}-
D\frac{\prt^2G}{\prt y^2}=0 
\ee
with the initial conditions $ G(0,x,y)=\delta(x)\delta(y)$ and the 
boundary conditions on the infinities $ 
G(t,\pm\infty,\pm\infty)=0$
and the same for the first derivatives respect to $x$.
Performing the Laplace and the Fourier transforms with respect to time and 
the $x$--coordinate, correspondingly, one obtains the solution for the 
subdiffusion along the structure axis \cite{baskin1}
\be\label{comb2}
G(t,x,0)=\frac{D^{1/2}}{2\pi\sqrt{\tilde{D}t^3}}\int d\tau
\exp\left[-\frac{x^2}{4\tilde{D}\tau}-\frac{D\tau^2}{t}\right] \, .
\ee
The total number of particles on the structure axis decreases with time
\be\label{c_1}
\langle G\rangle=\int_{-\infty}^{\infty}G(t,x,0)dx=
\frac{1}{2\sqrt{\pi D t}}\, .
\ee
Therefore this solution of Eq. (\ref{comb2}) describes the subdiffusion 
when the number of particles is not conserved. In what follows 
consideration, this point will be bearing in mind \cite{add1}. Therefore,
this solution corresponds to the subdiffusion with the second moment 
or the mean squared displacement along the structure axis of the form
\be\label{comb3}
\langle x^2(t)\rangle=
\frac{\langle x^2(t)G(t,x,0)\rangle}{\langle G(t,x,0)\rangle}=
\tilde{D}\left(\frac{\pi t}{D}\right)^{1/2} \, .
\ee
First, let us consider a transport problem on the comb structure due to 
the constant convection velocity $v_x=v_0\delta(y)$. Here and in what 
follows one bears in mind that the dimension of $v_sx^s$ is 
$\tilde{D}x$. The diffusive term in (\ref{comb1}) is omitted, for simplicity. 
Then, we have
\be\label{comb4}
\frac{\prt G}{\prt t}+v_0\delta(y)\frac{\prt G}{\prt x}-D
\frac{\prt^2G}{\prt y^2}=0 \, .
\ee
Now applying the Laplace and Fourier transform, this equation is solved 
exactly with the solution in the form
\be\label{comb5}
G(t,x,0)=\frac{D^{1/2} x}{\sqrt{\pi t^3v_0^2}}
\exp(-Dx^2/v_0^2t) \, ,
\ee
and $G=0$ for $ x<0 $, since
the distribution function must be positive. Moreover, the is no transport
in the negative direction due to the initial conditions.
It corresponds to the normal diffusion with the second moment 
\be\label{comb6}
\langle x^2(t)\rangle=\frac{v_0^2}{D} t \, .
\ee
Therefore, to obtain superdiffusion on the comb structure, it is 
not enough  to add a fast process {\em e.g.} due to the constant 
convection.
In what follows we discuss some mechanism for the superdiffusion,
where we suggest an inhomogeneous convection.

A superdiffusive L\'evy--like process corresponds to infinite flights
during a finite fixed time. Such a process could be a realization
of an inhomogeneous current in the $x$-direction. For example,
the following two-dimensional convective current 
\be\label{adc_1}
{\bf j}=\left(v_sx^sG(t,x,y)\delta(y),-D\prt G(t,x,y)/\prt y\right)
\ee
after substituting in the Liouville equation (\ref{LE})
gives some modification of Eq. (\ref{comb4}) in the form
\be\label{adc_2}
\frac{\prt G}{\prt t}+\hat{V}_xG\delta(y)-D\frac{\prt^2G}{\prt y^2}=0 
\, ,
\ee
where $s>0$ and  
\be\label{adcc_1}
\hat{V}_xG =v_sx^s \frac{\prt G}{\prt x}+sv_sx^{s-1}G \, .
\ee
After the Laplace transform with respect to the time, one presents the 
solution in the form 
\be\label{comb7}
G(p,x,y)=f(p,x)\exp\left[-(p/D)^{1/2}|y|\right] \, ,
\ee
where we used that 
\[\frac{\prt^2}{\prt 
y^2}\exp\left[-\sqrt{\frac{p}{D}}|y|\right]=
\left[\frac{p}{D}-2\sqrt{\frac{p}{D}}\delta(y)\right]\exp\left[-
\sqrt{\frac{p}{D}}|y|\right]  .\]
Substituting Eq. (\ref{comb7}) in Eq. (\ref{adc_2}),  
 we obtain then the following equation for $f\equiv f(p,x)$
\be\label{comb8}
v_sx^sf^{\prime}+sv_sx^{s-1}f+2[pD]^{1/2}f=\delta(x) 
\ee
with the following boundary conditions 
on the infinity  $f(p,\infty)= 0$ and $f(p,x)=0$ for $x<0$.
For $s<1$, the solution of Eq. (\ref{comb8}) has the following form
\be\label{comb8_a}
f=\frac{\Theta(x)}{v_sx^s}
\exp\left[-\frac{2(Dp)^{1/2}x^{1-s}}{v_s(1-s)}\right] \, ,
\ee
where $\Theta(x)$ is a step function: $\Theta(x)=1$ for $x\ge 0$ and it is 
zero for $x<0$.
It corresponds to some kind of superdiffusion along the structure axis 
where all moments are determined by the following distribution function
\be\label{adcc_2}
G(t,x,0)=\Theta(x)\frac{D^{\frac{1}{2}}x^{1-2s}}{v_s^2(1-s)\sqrt{\pi t^3}}
\exp\left[-\frac{Dx^{2-2s}}{v_s^2(1-s)^2t}\right] \, .
\ee
For example, $\lgl x^2(t)\rgl\propto t^{\frac{1}{1-s}}$.
When $s=0$, Eq. (\ref{adcc_2}) coincides with Eq. 
(\ref{comb5}). 
To avoid deficiency at $x=0$ where the solution is infinite, one needs to 
introduce a diffusion process, by means the second derivative with 
respect to $x$. In this case, the equations (\ref{adc_2}),(\ref{adcc_1}) 
and (\ref{comb8}) describe the space time evolution of the initial profile
of particles on the asymptotically large scale $ x\gg 1$.
Therefore, the solutions  (\ref{comb8_a}) and
(\ref{adcc_2}) are the asymptotic ones.
When $s\ge 1$, it is more convenient to look for a solution on a
class of generalized functions. The case with $s>1$ needs a special 
care that is deserved a separate consideration \cite{ib}. When 
$s=1$, it is a frontier case, with a strong L\'evy--like process.
The so--called log--normal distribution could be realized here as well.

We consider the  case of
inhomogeneous transport along the structure axis due to the inhomogeneous 
drift with $s=1$ and inhomogeneous diffusion of the form 
$D_{xx}(x,y)=\tilde{d}x^2\delta(y)$. To this end we add the diffusion 
$ -\tilde{d}x^2\delta(y)\prt G/\prt x$ to the $x$--component of the 
current in Eq. (\ref{adc_1}). The Fokker--Planck equation (\ref{comb1}) 
reads now 
\be\label{comb9_1}
\frac{\prt G}{\prt t}+\hat{V}_xG \delta(y) - D\frac{\prt^2G}{\prt y^2}=0 
\, ,
\ee
where diffusion along the structure axis is 
\be\label{comb9_2}
\hat{V}_xG=-\tilde{d}x^2\frac{\prt^2 G}{\prt x^2}-
(2\tilde{d}-v_1)x\frac{\prt G}{\prt x} +vG \, .
\ee
Again, performing the Laplace transform with respect to the time
and presenting the solution of Eqs. (\ref{comb9_1}) and (\ref{comb9_2})
in the form of Eq. (\ref{comb7}) we  obtain the following equation for 
$f\equiv f(p,x)$
\be\label{comb9}
-\tilde{d}x^2f^{\prime\prime}-(2\tilde{d}-v_1)xf^{\prime}+(2[pD]^{1/2}+v_1)f=
\delta(x) 
\ee
with the same boundary conditions as in Eq. (\ref{comb1}).
The following solution is obtained:
\be\label{comb10}
f(p,x)=\frac{\delta(x)}{2[pD]^{1/2}}+
\frac{C_1}{|x|^r}\, , 
\ee
where 
$r=\frac{1}{2}[(1-b)+\sqrt{(b+1)^2+8(dp)^{1/2}}] $,
and we denote  $b=v_1/\tilde{d}$ and $d=D/{\tilde{d}}^2$.
The modulus here indicates that the equation is invariant with respect to 
the inversion $x\rightarrow\, -x$.
We also used here that $-x\prt\delta(x)/\prt x=\delta(x)$ and 
$ x^2\prt^2\delta(x)/\prt x^2=2\delta(x) $ on a class
of constant probe functions. This $\delta$--function solution for the 
inhomogeneous equation can also be obtained by the Fourier transform.
In this case the specific solution in the Fourier space 
is $f_k=(2\sqrt{pD})^{-1}$, whose the inverse Fourier transform 
gives the first term in (\ref{comb10}). The constant $C_1$ could be 
specified from the initial conditions. It does not carry any important 
information for the large scale asymptotics, and, without loosing of the 
generality, it could be equaled to the unity: $C_1=1$.

To perform the inverse Laplace transform it is convenient to use
the long time asymptotics $t\rightarrow\infty$ ($p\rightarrow 0$).
In this case $r\approx 1+a\sqrt{p}$ with $a=2d^{1/2}/(b+1)$.
It is also convenient to present the second term in (\ref{comb10})
in the form of the exponential: $|x|^{-r}=e^{-r\ln|x|}$. Hence the inverse 
Laplace transform becomes a standard procedure \cite{korn}, and  we obtain 
the following asymptotic solution:
\be\label{comb11}
G(t,x,0)=\frac{\delta(x)}{2\sqrt{\pi D t}} +
\frac{a\ln|x|}{2|x|\sqrt{\pi t^3}}
\exp\big[-\frac{a^2\ln^2|x|}{4t}\big] \, .
\ee
We would like to admit here that $|x|\gg 1$ and the logarithmic 
function is positive.
The first term corresponds to the smoothly time--decaying  pining 
distribution,
while the second term, which is the most important for the asymptotic 
transport, corresponds to a some kind of the log-normal distribution. 
It should be underlined that the asymptotic solution of Eq. (\ref{comb11})
is independent of the coefficient $\tilde{d}$, which determines the 
diffusion along the axis structure. It is natural, since 
an asymptotic space--time  evolution of the initial distribution is 
determined by the drift component
of the current, but not by diffusion. It means that one could consider
the convective process analogous to Eqs. (\ref{comb4}) and (\ref{adc_2})
instead of the more general consideration in the framework of Eqs. 
(\ref{comb9_1}) and (\ref{comb9_2}). In this case, of course, $G(t,x,y)=0$ 
for $x<0$.

Let us calculate the second moment 
analogously to Eq. (\ref{comb3}). We have then
$\langle G\rangle=[1+4(\tilde{d}+v_1)]/\sqrt{\pi D t} $,
and the second moment is 
\[ \langle x^2 G\rangle=(8\sqrt{t}/a^2) \exp[4t/a^2]\, . \]
Finely, the displacement is
\be\label{comb14}
\sqrt{\langle x^2(t)\rangle}=\sqrt{\frac{\langle x^2G\rangle}{\langle 
G\rangle}}\propto e^{2t/a^2}\, ,
\ee
This exponentially fast spreading is a result of the L\'evy--like process
introduced inside the system by means of the inhomogeneous displacement
current.
There is another interesting interpretation
of the obtained result. It could be considered as the Poisson 
distribution respect to the squared logarithm. Namely, if we denote
$z=\ln^2 |x|$, then we have from (\ref{comb11}) 
\be\label{comb15}
P(z)dz=\frac{a}{2\sqrt{\pi t^3}}e^{-a^2z/4t}dz \, .
\ee
Hence, the average value of $z$ is
\be\label{comb16}
\langle z\rangle=\big[\int_0^{\infty}zP(z)dz\big]/
\big[\int_0^{\infty}P(z)dz\big]=4t/a^2 \, .
\ee
It could be understood as the fact that
time approaches to the infinity with the same rate as the squared 
logarithm of 
the $x$--coordinate, namely $4t\sim a^2\ln^2 |x|$. Therefore, returning to 
the log--normal part of the distribution function of Eq. (\ref{comb11})
we have, approximately, that its asymptotic behavior is
\be\label{comb17}
G(t\gg 1,|x|\gg 1,y=0)\sim \frac{1}{|x|\ln^2|x|} \, .
\ee
It is simply to see, that the flux is zero on the infinities.
Finely, one obtains that all  even moments of $x$ diverge on the 
large scale asymptotics. The essential difference between the 
distributions for subdiffusion of Eq. (\ref{comb2}) which gives for 
$|x|,t\gg 1,$ that $G(t,|x|,0)\propto \exp\left[-(x^4/t)^{1/3}\right]$, and 
the superdiffusion of Eq. 
(\ref{comb17}) is obvious. Therefore, the main mechanism of the transition 
from subdiffusion in Eq. (\ref{comb3}) to superdiffusion in Eq. 
(\ref{comb14})
on the comb structure is the inhomogeneous convection.

In conclusion, we would like to classify a possible response of a 
substrate with anomalous transport. As it is shown here it depends on the 
external forcing according to the power rate of the inhomogeneous 
convection $j_x(x,t)=v_sx^s\delta(y)G(t,x,0)$. As it follows from the 
solution (\ref{adcc_2}), when $s<0$ it is subdiffusion 
\cite{procaccia,klafter}.
As we shown here, when $s>0$ it is superdiffusion, Eq, (\ref{adcc_2}).
The homogeneous convection with $s=0$ corresponds formally to the normal 
diffusion as in Eq. (\ref{comb6}), but the effective diffusion coefficient
$v_0^2/D$ is determined by the external forcing $v_0$. The frontier case
with $s=1$ has two features. The first one corresponds to the log--normal
distribution of transport particles, where one deals not with a sum
of independent random variables but with their multiplication
\cite{shlezinger}. As the result the exponentially fast transport takes 
place (see Eq. (\ref{comb14})). The second feature is that the solution 
satisfies to the natural boundary conditions, when the flux of 
transporting particles equals zero on the infinities for any time.
When $s>1$, it corresponds to superdiffusion \cite{ib}, as well.

The final remark is that the relaxation along the structure axis is 
``inevitable'' process due to the item $\tilde{D}\delta(y)\prt G/\prt x$.
Therefore, anomalous transport is due to the convection flow only on
the asymptotically large scale $x\gg 1$ where the diffusion process 
could be omitted. This point is studied in detain in \cite{ib}, where
we shown that this approximation corresponds to Liouville--Green
asymptotic solution \cite{olver}.


\begin{thebibliography}{99}
\bibitem{add1} In fact, we can consider the total distribution function
$\tilde{G}(t,x)=\int dyG(t,x,y)$, which corresponds to the continuous time 
random walk model \cite{shlezinger} with the conservation of the total 
number of particles. Nevertheless, we prefer to consider $G(t,x,0)$, since
it is more relevant value for practical applications \cite{baskin2}.
\end{thebibliography}
\end{document}